\documentclass[referee,sn-standardnature]{sn-jnl}

\usepackage{amsmath}
\begin{document}

\title[High quantum efficiency parametric amplification via hybridized NLO]{High quantum efficiency parametric amplification via hybridized nonlinear optics}

\author*[1]{\fnm{Noah} \sur{Flemens}}\email{nrf33@cornell.edu}

\author[1]{\fnm{Dylan} \sur{Heberle}}

\author[1]{\fnm{Jiaoyang} \sur{Zheng}}

\author[1]{\fnm{Devin J.} \sur{Dean}}

\author[1]{\fnm{Connor} \sur{Davis}}

\author[2]{\fnm{Kevin} \sur{Zawilski}}

\author[2]{\fnm{Peter G.} \sur{Schunemann}}

\author*[1]{\fnm{Jeffrey} \sur{Moses}}\email{moses@cornell.edu}

\affil[1]{\orgdiv{School of Applied and Engineering Physics}, \orgname{Cornell University}, \orgaddress{\street{142 Sciences Drive}, \city{Ithaca}, \postcode{14853}, \state{New York}, \country{United States}}}

\affil[2]{\orgname{BAE Systems}, \orgaddress{\street{P.O. Box 868, MER15-1813}, \city{Nashua}, \postcode{03061}, \state{New Hampshire}, \country{United States}}}

\abstract{Parametric amplifiers have allowed breakthroughs in ultrafast, strong-field, and high-energy density laser science and are an essential tool for extending the frequency range of powerful emerging diode-pumped solid-state laser technology. However, their impact is limited by inherently low quantum efficiency due to nonuniform light extraction. Here we demonstrate a new type of parametric amplifier based on hybridized nonlinear optics. Hybridization of parametric amplification with idler second harmonic generation induces unusual evolution dynamics for a fully parametric amplifier – with saturating rather than cyclic gain – observed here for the first time. This allows highly uniform light extraction enabling unprecedented efficiency for a lossless amplifier with Gaussian-like intensity profiles – a 48-dB single-stage gain with 68\% quantum efficiency and 44\% pump-to-signal energy conversion – a several-fold improvement over the standard. Possessing both laser-like high quantum efficiency and the advantages of thermal-loading free parametric systems, this simple approach can be implemented widely and have significant impact by increasing several-fold the power available for science and industry.}

\maketitle

Technology based on optical parametric amplification (OPA) and its variants, developed over the past five decades \cite{akhmanov1965,Baumgartner79,DUBIETIS92,Cerullo03,Fattahi2014}, has contributed greatly to a 21st-century revolution in energetic picosecond and femtosecond pulsed sources of visible through mid-infrared light. With each advance followed breakthroughs in strong-field science, high-energy density science, and the nonlinear spectroscopy of ultrafast molecular and materials phenomena \cite{Baltuska02,Ledingham03,Nibbering05,Mourou2006,SALAMIN2006,Sansone2011,Danson15}. Unlike lasers, however, in which gain saturates and makes uniform extraction of pump energy feasible, the gain dynamics of OPA are cyclic. A perpetual back and forth energy exchange begins with a transfer of power from the pump to the desired signal and idler waves followed by a return of power to the pump \cite{Armstrong:62}. These dynamics are the root cause of low quantum efficiency (defined as the ratio of emitted signal photons to incident pump photons) in OPA, since they make uniform light extraction across the naturally bell-shaped transverse beam and pulse profiles of pump lasers impractical.

\begin{figure}
    \centering
		\includegraphics[width=4.5in]{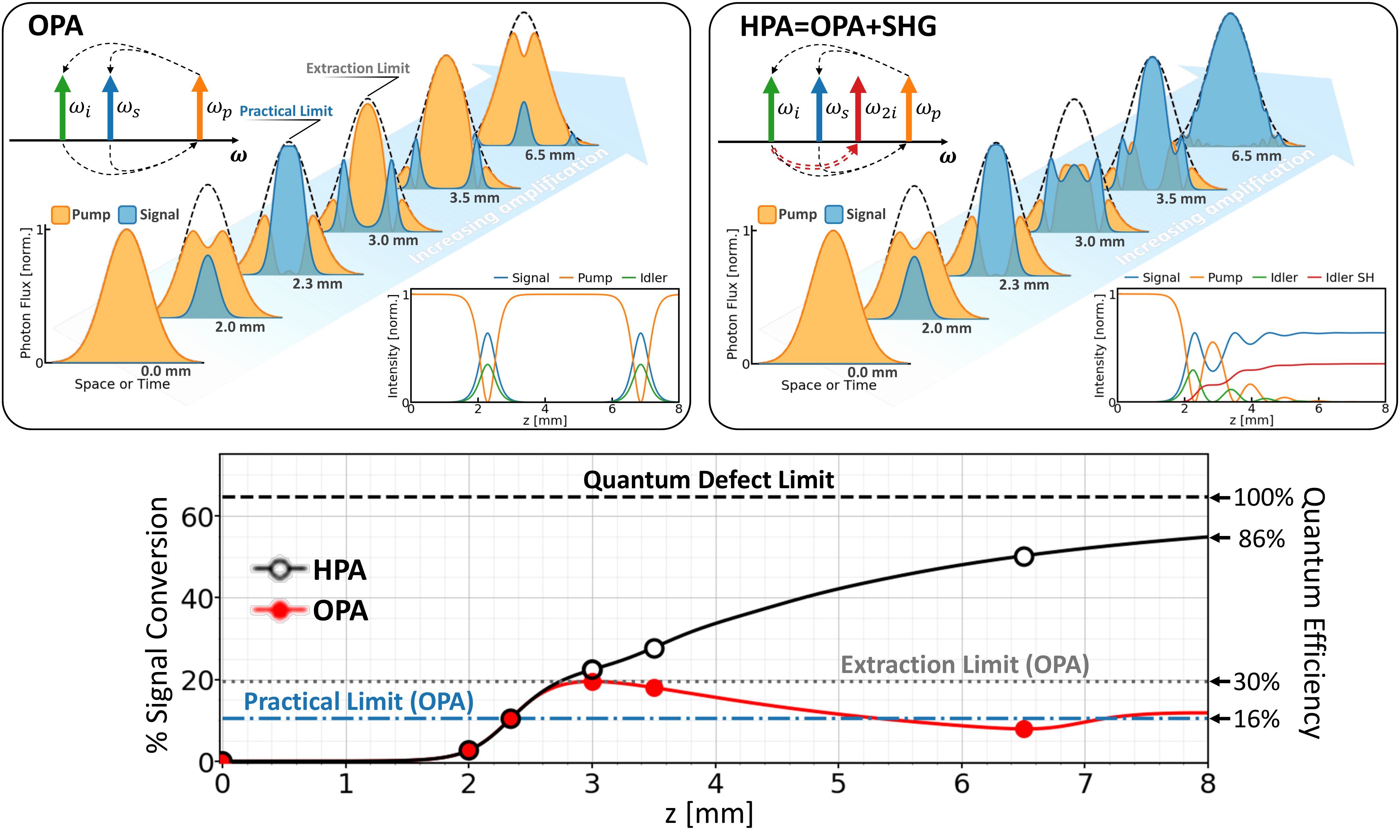}
        \caption{\textit{Top}: simulated pump-signal conversion dynamics of waves with Gaussian intensity profiles for OPA (\textit{left}) versus HPA (\textit{right}), representing a 1D slice across the center of a 3D spatiotemporal wave at specific propagation lengths. [Insets: photon exchange diagram and simulated (continuous-wave and plane-wave) wave propagation dynamics at the peak of the spatiotemporal profile.] In OPA, energy flow is cyclic, with cycle period dependent on the local pump and seed intensities. Temporary extraction of pump photons via conversion to the signal and idler begins at the profile center, working its way outwards and extraction is always highly nonuniform, resulting in low quantum efficiency integrated across the full 2+1D spatiotemporal extent. \textit{Bottom}: the evolving spatiotemporally integrated pump-to-signal energy conversion efficiency and quantum efficiency is shown for OPA and HPA, with circles corresponding to the specific lengths shown in the top panels. For OPA, in this example, the maximum energy conversion efficiency (quantum efficiency) peaks at 19\% (30\%) at 3.0 mm. This limitation to conversion efficiency is caused by non-uniform light extraction (\textit{`Extraction Limit (OPA)'}), and corresponds to a ring-shaped signal beam and spherical shell-shaped spatiotemporal profile, a shape that is usually impractical for applications. The maximum energy efficiency (quantum efficiency) achieved while maintaining a bell-shaped profile (\textit{`Practical Limit (OPA)'}) peaks at only 10\% (16\%) at 2.3 mm. In contrast, in HPA, energy flow has a quasi-saturating behavior characterized by damped oscillations, with oscillatory convergence of the pump field to the signal and idler SH fields. The saturating gain allows nearly uniform pump light extraction that asymptotically approaches 100\% quantum efficiency (\textit{`Quantum Defect Limit'}). For example, at 6.5 mm, $>$50\% energy conversion ($>$78\% quantum efficiency) is achieved, or $\sim$5 times greater efficiency than the OPA practical limit. At this length, the residual pump wave is characterized by low-amplitude ripples that are most pronounced towards the profile edges, and the amplified signal beam closely matches the profile of the incident pump. In all simulations, the initial pump/seed pulse energy ratio is 10$^5$ and the quantum defect, $\omega_s/\omega_p$, is 0.64. All simulation parameters are common for OPA and HPA cases, the only difference being the condition of phase-matched idler SHG included for HPA.}
    \label{fig:concept}
\end{figure}

Figure \ref{fig:concept}, \textit{top left}, illustrates this well-known problem. Since the OPA gain cycle period lengthens with decreasing local pump intensity, pump energy is only temporarily extracted at each location starting inside and moving outwards along the transverse profile. The limit to signal conversion efficiency imposed by this non-uniform pump extraction is several times smaller than the limit due to the quantum defect, i.e. the ratio of signal and pump photon energies (Fig. \ref{fig:concept}, \textit{bottom}). Deepening the problem, maximum extraction occurs at an amplifier length at which significant back-conversion of signal power at the beam center results in a ring-shaped beam. If a bell-shaped signal beam is desired, as is usually the case for applications, the practical conversion limit can be up to another factor of two smaller. Generally, for Gaussian spatiotemporal intensity profiles, the quantum efficiency ranges from 10-30\% depending on the amplifier's gain. The pump-to-signal energy conversion efficiency -- further reduced by the size of the quantum defect -- typically ranges from single-percent to $\sim$20\%. 

Low quantum efficiency means the great majority of pump photons go unused. This inefficiency is of particular significance today with the advent of record-high average power femtosecond and picosecond diode-pumped solid-state (DPSS) lasers \cite{Jauregui2013,Fattahi2014,Sincore18,Saraceno2019}, recently exceeding ten kilowatts \cite{Muller:20}, that can be used to pump OPA. It remains a challenge to \textit{efficiently} convert that power to ultrafast sources across the visible and infrared spectrum where they can enable new and more cost-effective applications for science, medicine, and industry, and in the mid-infrared transmission windows to advance ranging and sensing technologies. Our present work addresses this barrier for laser science and applications resulting from poor OPA efficiency. We demonstrate a way to fundamentally alter the nonlinear dynamics of OPA to enable uniform pump energy extraction and thus, high quantum efficiency, via a hybridization of three-wave mixing processes in an ordinary OPA crystal. The result is a first solution that possesses both high quantum efficiency through laser-like spatiotemporally uniform pump extraction and the advantages of an OPA such as the absence of thermal loading, which is essential for scaling OPA with the high average powers of emerging DPSS pump lasers.

Historically, several methods for boosting pump light extraction have been demonstrated in bulk-crystal based parametric amplifiers, though none are widely employed. 74\% net quantum efficiency in a sub-Joule scale, nanosecond chirped pulse power amplifier with 10 dB gain has been achieved via top-hat beam and pulse shaping \cite{Bagnoud:05}. This is the simplest use of conformal profiles \cite{Begishev:90,Moses:11,Cao:18,Mackonis:20,Fischer:21}, a general technique for synchronizing OPA conversion cycles across the spatiotemporal transverse extent of the interacting waves by pre-amplifier beam and pulse shaping. However, the difficulty in making the shaping process itself efficient, especially for femtosecond and picosecond pulses, has limited the use of this practice. 

Alternatively, it has been shown that idler loss during amplification can inhibit the sum-frequency generation process responsible for back-conversion of the amplified signal wave to the pump, thereby allowing more uniform extraction of photons from the pump. \textit{Jeys} demonstrated an eight-pass parametric generator with the idler absorbed between passes by a roof prism \cite{Jeys:96}. 80\% pump photon depletion was achieved, though the net quantum efficiency was limited to 45\% due to significant optical reflective losses. More recently, several works have explored concurrent idler loss during amplification by means of material absorption or radiation \cite{Ma:15, El-Ganainy:15, Zhong:16, Ma:17}. The OPA evolution dynamics in the presence of idler loss were found to exhibit damped power oscillations with a gradual increase in pump-to-signal energy conversion efficiency, an indication of the system's non-conservative (non-parametric) nature. These quasi-saturating gain dynamics allowed vastly improved pump extraction, with an impressive 70\% internal quantum efficiency achieved with $\sim$20 dB gain in a YCOB crystal doped with Sm$^{3+}$ to induce idler absorption \cite{Ma:15}. An example of the `gain via loss' paradigm \cite{Perego2018,Feng2017}, improved conversion efficiency comes at the expense of loss of the idler power as well as the resulting thermal loading. Additionally, the approach requires the engineering of new, application-appropriate lossy nonlinear crystals, limiting its applicability. 

\subsection*{Hybridized parametric amplification}\label{HPA}

Recently, several of us proposed a hybridized parametric amplification (HPA) scheme that employs concurrent idler self-frequency doubling to alter the nonlinear dynamics of OPA. It was discovered that displacement of idler photons by second harmonic generation (SHG) to a coherent copropagating idler second harmonic (SH) wave shares a general dynamical form with the case of OPA undergoing concurrent idler linear absorption: that of a damped nonlinear (Duffing) oscillator \cite{Flemens:21}. These dynamics allow uniform pump energy extraction from spatiotemporally bell-shaped beams (Fig. \ref{fig:concept}, \textit{top right}), approximating the gain saturation that is inherent to a laser amplifier but keeping advantages unique to fully parametric (lossless) amplification, such as phase-matching limited gain band tunability and no thermal loading. Capitalizing on this behavior, our numerics predict that HPA allows pump-to-signal conversion efficiency approaching full quantum efficiency -- exceeding the practical OPA limit by severalfold (Fig. \ref{fig:concept}, \textit{bottom}) -- while producing an output beam profile closely matching that of the bell-shaped pump, and without any transfer of power to the gain medium. This may be a key enabling feature for high-average-power and high-energy applications that could enable new breakthroughs in ultrafast laser science. In the following, we describe an experimental demonstration of this principle in a single-stage, 48-dB mid-infrared amplifier with 68\% quantum efficiency and in an ordinary OPA bulk crystal, with efficiency far exceeding its ordinary OPA counterpart. Moreover, we clearly observe the novel evolution dynamics enabling the improved performance, with experimental data closely matching a theoretical spatiotemporal propagation analysis.

\subsection*{Experimental design and setup}\label{Setup}

\begin{figure}
    \centering
		\includegraphics[width=\textwidth]{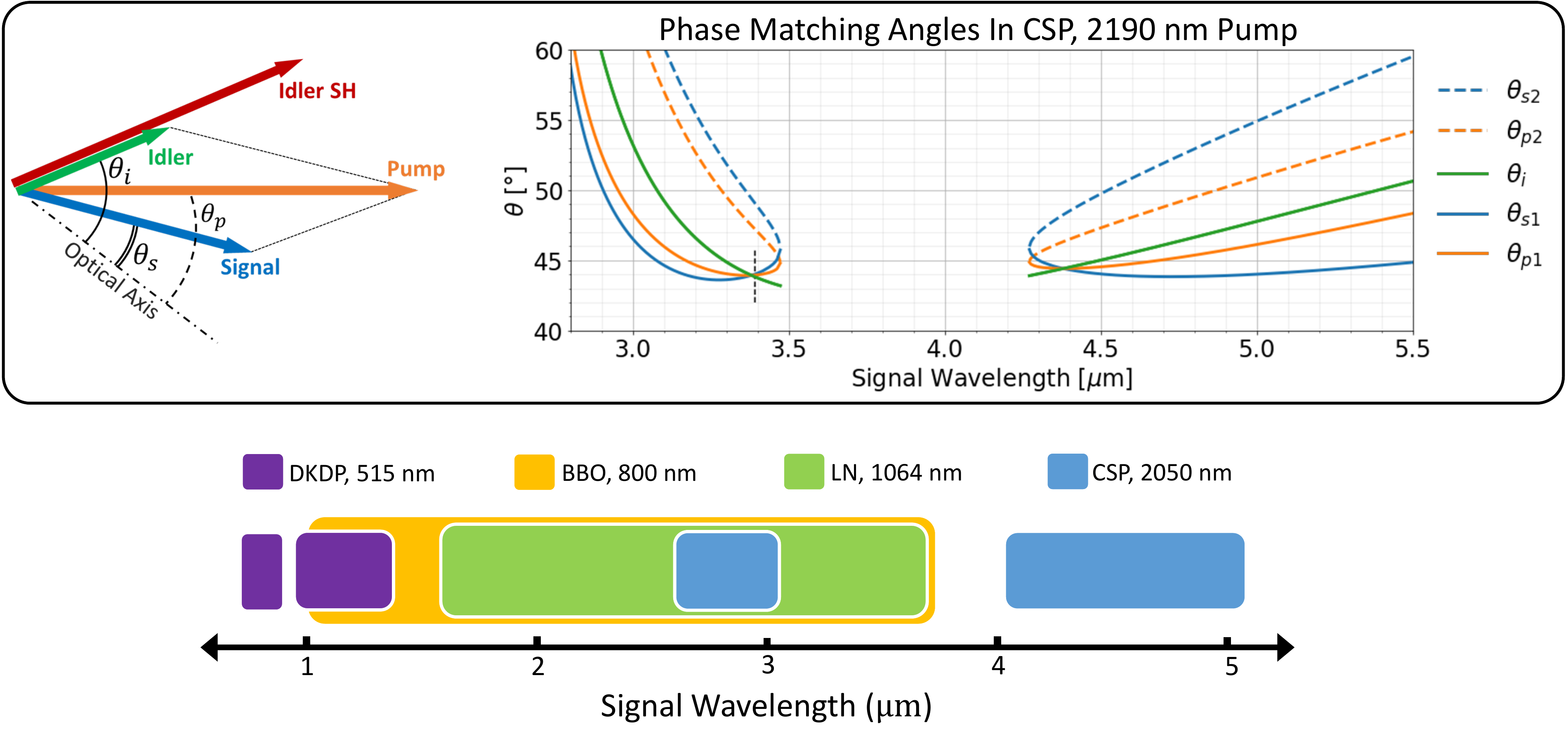}
        \caption{\textit{Top}: Joint Type-I OPA and Type-I idler SHG birefringent phase-matching diagram (showing the angles of wave-vectors relative to a uniaxial crystal's optical axis) and calculated angles in CSP at the experimental pump wavelength of 2190 nm. Generally, two non-collinear configurations are possible across the signal tuning range, indicated by $(\theta_{p,1},\theta_{s,1})$ and $(\theta_{p,2},\theta_{s,2})$, which share a common $\theta_i$ that is fixed at the SHG phase-matching angle. A wide signal wavelength tuning range exists below 3.5 \textmu m and above 4.2 \textmu m with non-collinear angles of up to a few degrees. At the experimental signal wavelength of 3390 nm, a solution exists with near-collinear wave-vector geometry. The \textit{bottom} panel illustrates a wide signal tuning range for other combinations of common bulk OPA media and common DPSS pump laser wavelengths. Ranges indicate joint Type-I OPA and Type-I idler SHG phase matching for pump-signal angles limited to 4 $\deg$.}
    \label{fig:NCPM}
\end{figure}

For this proof-of-principle demonstration, we chose a signal wavelength relevant for applications in the mid-wave infrared atmospheric passband at 3390 nm, and a pump wavelength of 2190 nm, within or not far from holmium-, thulium-, and chromium-doped DPSS ultrafast laser emission bands. For this wavelength range, CdSiP\textsubscript{2} (CSP) is a convenient OPA crystal, with high nonlinear coefficient, 2.45-eV band edge and good transparency out to 6.5 $\mu$m, and favorable Type-I OPA phase-matching when pumped near 2 $\mu$m \cite{Sanchez:14,Liang:17}. A previous analysis of non-collinear birefringent phase-matching for simultaneous Type-I OPA and Type-I idler SHG found broad applicability to common OPA crystals and pump wavelengths, with wide signal tunability  \cite{Dean2021Apr}. Fig. \ref{fig:NCPM}, \textit{top} extends this analysis to CSP at our choice of pump wavelength, and finds a range of signal wavelengths over which the joint phase-matching angles can be satisfied. At a signal wavelength of 3390 nm, the phase-matching conditions support a near-collinear wave-vector geometry ($\theta_p, \theta_s,  \theta_i \simeq 44 \deg$), though we note that Poynting-vector non-collinearity and therefore spatial walkoff occurs due to the involvement of two waves with extraordinary polarization (pump and idler SH). Fig. \ref{fig:NCPM}, \textit{bottom} illustrates that this Type-I birefringent phase-matching scheme is more broadly applicable to common OPA crystals and common DPSS pump laser wavelengths and allows tuning over a wide signal wavelength range with small non-collinear angles. 

Experimental implementation of HPA requires no significant modifications to the simple apparatus of an OPA device. The only additional consideration in our experiment was ensuring that all idler frequencies are phase matched for SHG. This was achieved by insertion of a spectral bandpass filter in the signal seed path. To satisfy the phase-matching condition at near-collinear angles, we used a Rochon prism to spatially combine the orthogonally polarized pump (up to 100 $\mu$J/pulse at 10 kHz pulse rate) and signal seed (0.7 nJ/pulse) beams before passing through a 6-mm CSP crystal. The beam sizes ($\sim$1-mm 1/e$^2$ diameter) and pulse durations ($\sim$1-ps FWHM, slightly chirped, transform-limited duration of $\sim$600 fs) were chosen to approximately match, ensuring the full spatiotemporal extent of the pump was overlapped with signal light. Within the device, transfer of pump photons to the signal and idler is accompanied by frequency doubling of the idler (at 6186 nm) to its SH (at 3093 nm). We monitored pump, signal, and idler SH beams at the device exit as the pump power was increased to its maximum. Since the idler SHG process is efficient and the exit face of the CSP crystal was anti-reflection (AR) coated for 2.0-4.8 $\mu$m, monitoring the low-power residual idler beam was impractical. Measured spectra are shown in Extended Data Fig. \ref{fig:spectra} alongside corresponding simulated spectra and temporal profiles.

\begin{figure}
    \centering
		\includegraphics[width=\textwidth]{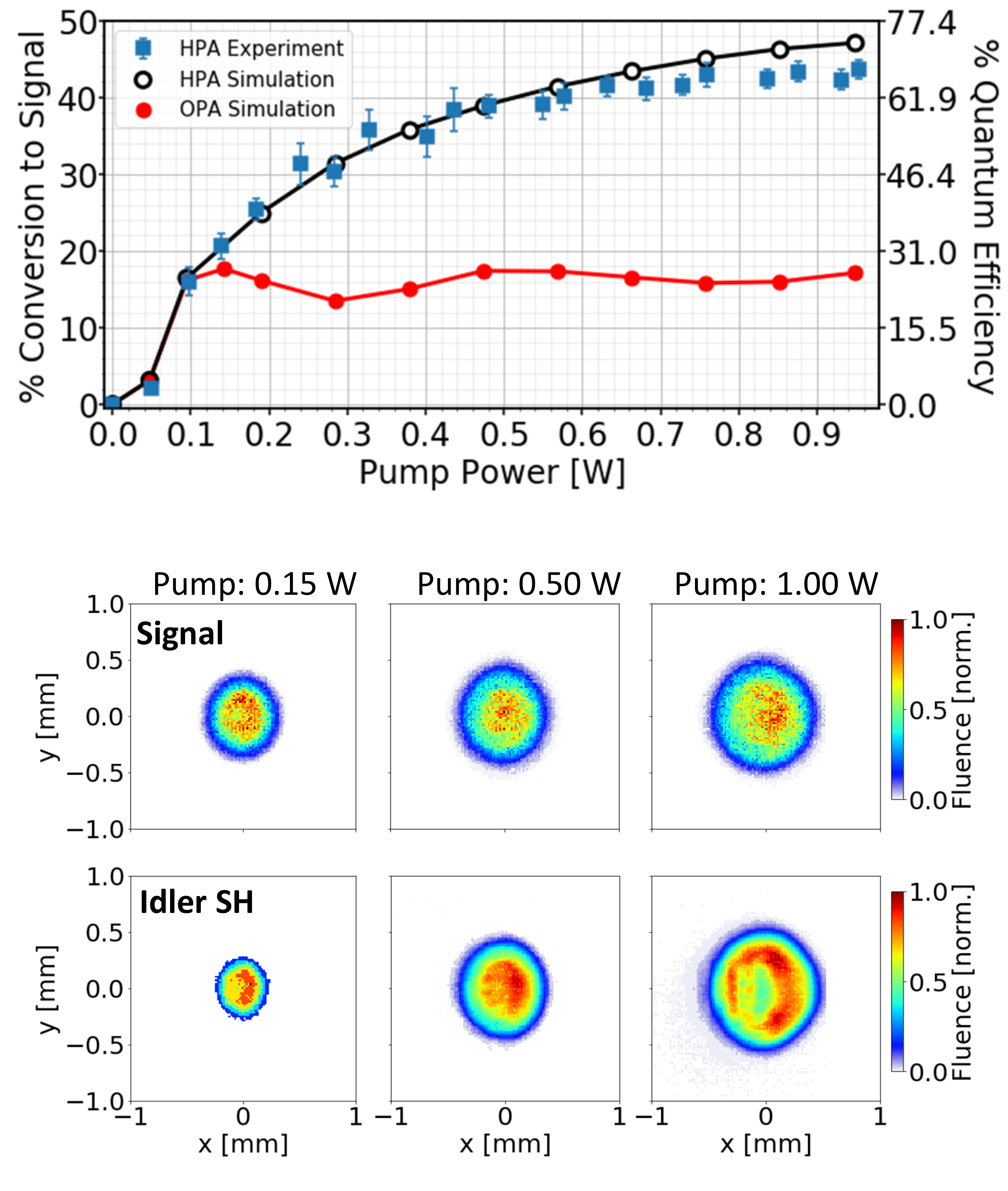}
        \caption{\textit{Top}: pump-to-signal energy conversion efficiency comparing the HPA experiment (blue squares) to simulated HPA (black circles) and the equivalent OPA without SHG (red circles). Error bars indicate the standard deviation of signal powers measured by a thermopile-based power meter over 1 minute. \textit{Bottom}: measured signal and idler SH beam profiles with increasing applied pump power.}
    \label{fig:experimentalData}
\end{figure}

\section*{Results}\label{Results}

\subsection*{High quantum efficiency via uniform light extraction}\label{HighEfficiency}

The pump-to-signal energy conversion efficiency internal to the device as a function of input pump power is shown in Fig. \ref{fig:experimentalData}, \textit{top}, along with the results of spatiotemporal wave propagation simulations using the model and framework of \cite{Flemens:21}. The experimentally measured (blue squares) and simulated (black circles) data for the HPA process are in good agreement, with a clear trend of monotonically increasing efficiency. This behavior is as expected given the dynamics of HPA, which inhibit signal back-conversion and provide saturating gain at all spatiotemporal coordinates of the amplifier. Fig. \ref{fig:experimentalData}, \textit{bottom} corroborates this behavior, illustrating a signal beam profile (imaged at the crystal exit plane) that expands in size with increasing pump power, becoming roughly equal in size to the incident pump beam as the pump light extraction becomes more uniform. The slightly lower measured performance at higher powers of the HPA system compared to simulations might be explained by the observed onset of idler SHG reversal at the beam center in the presence of spatial walk-off (an effect not captured by the simulations). Idler SHG reversal is a consequence of imperfect phase matching across the full transverse wave-vector and spectral content of the fields. Since for higher pump powers, idler SHG begins earlier in the crystal, decoherence due to phase mismatch is more pronounced. In other words, idler SHG decoherence limits the device length, and thus greater efficiency requires shorter length and higher pump intensity. In our experiment, an observed spatial asymmetry of the idler SH indicates significant Poynting vector walk-off (Fig. \ref{fig:experimentalData} \textit{bottom}), which would have a more pronounced effect for smaller beam size. Thus, our choices of beam size, pump intensity, and crystal length represents an optimal balance between decoherence and spatial walk-off effects. 

The maximum internal pump-to-signal energy conversion efficiency achieved is 44\%, corresponding to 41 \textmu J of measured energy per signal pulse at 100 \textmu J incident pump pulse energy due to imperfect AR coatings (4\% reflections at each face), and with 18 \textmu J of generated idler SH. These results correspond to an internal quantum efficiency of 68\%, corroborated by a separate analysis of the measured depletion of the pump power (see \textit{Supplementary Section 1}), and a 48-dB signal gain. This represents a new standard for lossless, thermal-loading free parametric amplification, based on the novel underlying nonlinear dynamics of the HPA approach. The efficiency far exceeds that of standard OPA (Fig. \ref{fig:experimentalData}, \textit{red circles}), which peaks at 18\% at 140 mW of pump power with a donut-shaped beam due to significant back-conversion at the beam center. Achieving a comparable amount of signal energy as our HPA result in a single OPA stage while maintaining a bell-shaped amplified beam shape would require 4.5 times the pump power and would only be 9\% efficient, according to simulation. 

\begin{figure}
    \centering
		\includegraphics[width=\textwidth]{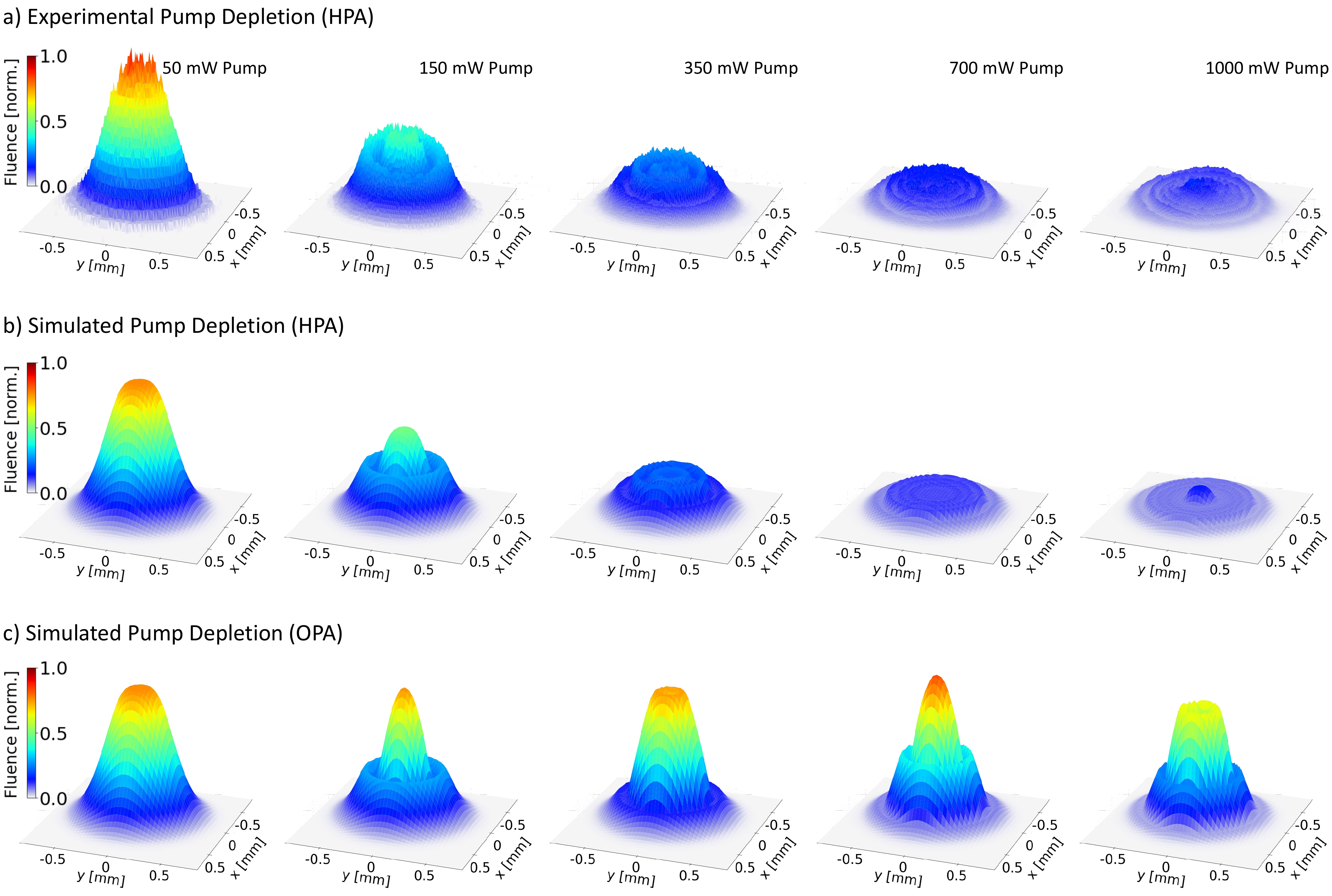}
        \caption{(a) Depleted experimental pump beam profiles imaged to CSP crystal end facet and normalized to input pump energy, and corresponding normalized simulated output pump beam profiles from (b) HPA and (c) OPA. [Full datasets for (a) and (b) are provided in Supplementary Video 1.]}
    \label{fig:pumpDepletion}
\end{figure}

\subsection*{Observation of novel nonlinear dynamics}\label{ObservationOfDynamics}

By examining the residual pump beam profile as a function of incident pump power, we can also clearly discern the novel nonlinear dynamics that allow uniform pump extraction in HPA. Fig. \ref{fig:pumpDepletion}a shows the measured residual pump beam at the output of the CSP device normalized in each frame to the input pump power. Depletion starts from the center of the pump beam and moves outward until a significant amount of energy is extracted by the signal beam. As the \textit{damped} oscillatory dynamics of HPA occur at each local transverse coordinate but with asynchronous oscillation period, as predicted in Fig. \ref{fig:concept}, a concentric ring pattern is formed across the beam that diminishes with increasing pump power. The measured profiles exhibit a remarkable match to the HPA simulations (Fig \ref{fig:pumpDepletion}b), and highly contrast with simulations of the equivalent OPA device (Fig. \ref{fig:pumpDepletion}c). As seen, the \textit{undamped} oscillations in pump extraction characteristic of OPA also lead to the formation of concentric rings, but these do not diminish with increasing pump power, preventing uniform extraction across the pump beam at any power and resulting in low quantum efficiency. Thus, we clearly observe the hybridized nonlinear optical system dynamics leading to uniform pump extraction predicted in \cite{Flemens:21} for the first time, and confirm the highly contrasting pump extraction dynamics of HPA and OPA. 

\section*{Discussion}\label{Discussion}

Since the HPA implementation demonstrated here is no less practical to implement than standard OPA, working for unshaped beams, without idler loss and thermal loading, and with standard bulk crystals, the technique offers a solution that could be easily and widely adopted for the efficient frequency down-conversion of modern high-power DPSS lasers. Simultaneous OPA and idler SHG Type-I birefringent phase matching will generally be possible across common OPA crystals and DPSS laser wavelengths for pulses hundreds of femtoseconds and longer in duration, making it relevant to amplifiers based on Yb-doped, Ho-doped, and many other gain media. This means HPA could be employed to efficiently extend ultrafast lasers that can produce high average power and/or high pulse energy to the large spectral ranges where suitable gain media do not exist. For many applications, HPA may prove to be even more simple and practical to implement than OPA, as it allows high efficiency to be achieved even at high gain (48 dB in this demonstration). Thus, HPA can also obviate the common practice of dividing available pump power into high-gain pre-amplifier stages and a low-gain power amplifier. This technique is employed to marginally improve overall efficiency since the practical OPA efficiency limit increases with lower gain. Elimination of the need for two or more amplifier stages cuts losses and could prove to be an important simplification to parametric amplifier architecture, allowing technologies that are more robust, portable and cost-effective.

Another possible significant advantage of HPA is illustrated by Fig. \ref{fig:experimentalData}, \textit{top}, which shows a significant noise reduction as the amplifier is pushed into the gain saturation regime at high pump powers. In OPA, gain non-uniformly saturates and reverses to loss at the practical OPA limit, thus preventing a strong reduction in noise at peak efficiency and causing increased fluctuations in transverse profiles (as well as in the power spectrum in optical parametric chirped pulse amplification (OPCPA)) \cite{Manzoni:11, Feng:21}. In contrast, in HPA, gain saturation takes place at all transverse coordinates simultaneously. This and the suppression of back-conversion produces an observed clamping effect on noise at high conversion efficiency as the signal transverse profile asymptotically grows, approaching the form of the incident pump. This may also help to increase the pulse contrast, as recently predicted for OPA with linearly absorbed idler \cite{Hu2020}. 

The amplified signal bandwidth obtained at full efficiency corresponds to a 340-fs transform limited pulse duration. The simple bulk-crystal birefringent phase matching approach taken here is not likely to offer, for any given application, a few-cycle phase-matching bandwidth for both OPA and SHG processes simultaneously in available crystals. However, we propose several plausible routes to high-quantum-efficiency few-cycle pulse amplification via HPA. For example, HPA could potentially accommodate both few-cycle pump and signal pulses via a dual-chirp scheme with matching pump and signal chirp rates. Such a configuration, employed previously for spectral compression of the idler in difference frequency generation \cite{Luo:06}, can sharply reduce the bandwidth of the idler pulse. A hybridized parametric amplifier using this configuration would then require a broad phase-matching bandwidth for the OPA process only since the idler SHG process would be narrowband, thus allowing compatibility with standard noncollinear OPA phase-matching angles employed for few-cycle amplification \cite{Cerullo03}. This approach is naturally suited to femtosecond pump lasers such as Ti:sapphire amplifiers or DPSS lasers with hollow-core fiber post compression. Fourier-domain OPA \cite{Schmidt2014}, in which an angularly dispersed signal overlaps with a narrowband pump pulse at the image plane of a diffraction grating, could potentially accommodate HPA in few-cycle amplifiers that employ a relatively narrowband pump pulse, since locally both the pump and signal are spectrally narrow in the amplifier medium. Fourier-domain HPA could therefore potentially serve as a high-efficiency alternative to few-cycle OPCPA pumped by picosecond lasers with the Fourier-domain OPA capacity for allowing spatially engineered nonlinear crystals at the Fourier plane \cite{Leblanc:20}, a route to the possible accommodation of an octave-spanning signal bandwidth. Quasi-phase matching can be used to suit the needs of many potential applications that do not require large crystal apertures, using superlattice and other poling approaches \cite{Chou:99,Liu:02,Norton:03,Lifshitz:05,Flemens:21}.

An especially intriguing possibility is to make use of the significant energy displaced to the co-propagating idler SH field to pump a subsequent signal amplification stage, which is practical when the idler SH frequency is greater than the signal frequency. This was proposed in \cite{Flemens:21} as a potential route for overcoming the quantum defect limit of OPA, i.e., for producing a double-stage quantum efficiency above 100\%, meaning more than one signal photon is produced for every incident pump photon. Cascading HPA stages in this way could have significant impact for atmospheric laser science applications by improving the efficiency of long-wave infrared generation pumped by near-infrared lasers, where the quantum defect is on the order of ten percent and typical OPA energy efficiency is at the single to few percent level, making it impractical for many applications.

While OPA has had a tremendous impact on ultrafast laser science and technologies, it has perennially suffered from low quantum efficiency. The new concept of parametric amplification employing hybridized nonlinear optics overcomes this issue with a novel combination of positive attributes found in ultrafast lasers and parametric amplifiers, including gain saturation without back-conversion and thus uniform light extraction even at high gain, a tunable gain band, and an absence of thermal loading. Through a capacity for high efficiency at a high single-pass gain, it can also be used to simplify OPA-based architecture. HPA thus sets a new quantum efficiency standard for parametric amplification while possessing the attributes required to handle high average power and broadly extend the frequency range of emerging DPSS ultrafast pump laser technology.

\backmatter

\section*{Methods}\label{Methods}
\subsection*{Pump and Seed Generation}
The 2190 nm HPA pump and 3390 nm HPA seed sources were each derived from a common home-built noncollinear 10-fs near-IR OPCPA system spanning 680-950 nm. The OPCPA is white-light seeded and pumped by a 1.03 \textmu m, 10-kHz, commercial picosecond Yb:YAG amplifier (Amphos) frequency doubled to 515 nm. To generate the mid-IR wavelengths used for this experiment, we employed adiabatic difference frequency converters (ADFC) based on aperiodically poled lithium niobate, similar to that used in Ref. \cite{Krogen2017Apr}, designed by our group and manufactured by HC Photonics. For the pump beam line, we used a 4-f pulse shaper (Phasetech) to select  a 0.5-nm band at 700.8 nm from the near-IR OPCPA front end, which was subsequently converted to 2190 nm in a 2-cm LiNbO\textsubscript{3} ADFC pumped by the 1.03-\textmu m Yb:YAG amplifier. This light was then amplified by OPA to 100 \textmu J (1 W average power) using a 1030-nm pumped 2-mm periodically poled LiNbO\textsubscript{3} crystal (HC Photonics) pre-amplifier stage and 3-mm BiBO (Castech) power amplifier stage. The OPA pump source is tunable via selection of the near-IR band via the pulse shaper. For the seed beam line, we used a bandpass filter (Andover) to select a fixed 790 $\pm$ 0.5 nm band from the near-IR OPCPA front end, which was converted to 0.7-nJ, 3.39-\textmu m pulses in a parallel ADFC stage.

\subsection*{HPA Stage}
The HPA experiment was performed using a 6-mm CSP crystal grown by BAE Systems. The input and output faces of the crystal were AR coated (TwinStar) for 2.0-4.8 \textmu m with 4\% reflection losses on each face measured for the pump and signal wavelengths. The pump beam path incorporated a translation stage to fine tune the temporal overlap of the beams. A Rochon prism (RPM10, Thorlabs), which converts incident orthogonally polarized beams to a collinear output, was used to achieve spatial overlap of the pump and seed beams. The Rochon prism additionally served as a control for pump power through tuning of an upstream half-wave plate. The beams were then reduced in size to approximately 1-mm 1/e$^2$ diameter and recollimated in a 4x shrinking magnification telescope. Two steering mirrors after the telescope were used to align the common beam path into the CSP crystal.

\subsection*{Diagnostics}
The input pump and seed energies were measured with a thermopile average power and energy sensor (Ophir 3A). At the output of the CSP crystal, reflections of the beams from a CaF\textsubscript{2} beam sampler were imaged with a CaF\textsubscript{2} lens in a 2F configuration to a microbolometer camera (DataRay WinCamD-IR-BB). A Wollaston prism was used to spatially separate the beams on the camera by polarization. The transmission through the beam sampler was used to measure the power of each beam with the thermopile sensor. Spectral filters (Thorlabs 2250$\pm$500 nm, Thorlabs 3500$\pm$500 nm, and Andover 3050$\pm$100 nm) were used to isolate each beam for spatial profile and power measurements. The crystal angle, signal alignment, and pump timing were adjusted to simultaneously optimize the beam shape and power of the signal beam. For the signal conversion efficiency measurements, the beam sampler was removed and the 3500$\pm$500 nm filter was used to isolate the signal.

\subsection*{Numerical Model}
The 2D spatial plus 1D temporal HPA and OPA simulation results reported in Figs. \ref{fig:experimentalData} and \ref{fig:pumpDepletion} employed a 1D temporal grid numerically propagated along the crystal axis using a Fourier split-step approach that captured the full dispersion series of CSP (using Sellmeier equations reported in \cite{ZAWILSKI20101127}) and a unidirectional propagation model of nonlinear wave mixing (as described in \cite{Flemens:21}) between signal, idler, idler SH, and pump fields by quadratic-order electric polarizability. This captured OPA between pump, signal and idler fields, and frequency doubling of the idler field to the idler SH. In addition, to capture possible parasitic wave mixing, we included two additional fields and nonlinear processes: SHG of the signal to a signal SH field, and OPA between signal and idler SH to a field at their difference frequency. To capture 3D spatiotemoporal dynamics, we defined the initial transverse field profiles of the pump and signal to be perfect Gaussians in space and time. All other fields were initialized to zero. Simulations captured collinear propagation through a CSP crystal with theta angle of 43.9 degrees. 

Fields were sampled in space by a 62 by 62 grid, and the temporal pulse propagation dynamics for each spatial grid point were computed individually by the Fourier split-step method described above. Thus, linear temporal effects (including group-velocity walk-off, group-velocity dispersion, and higher order dispersion) were accounted for, but diffraction and spatial walk-off were not. This was justified by separately carried out simulations including diffraction and spatial walk-off that indicated only minor features due to these effects, and which was corroborated by the excellent match between experimental and numerical data found in our study, as shown in Figs. \ref{fig:experimentalData}, \ref{fig:pumpDepletion}, and Supplementary Fig. 1. As noted, however, small discrepancies between experimental and numerical data are likely due to minor Poynting-vector walk-off of the extraordinary-polarized
pump and idler SH beams from the ordinary-polarized signal and idler beams, which was not captured in the simulations. The pump and signal durations were chosen based on the filter bandwidths in the experiment resulting in transform limited pulse durations of 1.2 ps for the 2.2 \textmu m pump and 0.9 ps for the 3.39 \textmu m signal. The pump and signal beam diameters were 1.0 and 1.2 mm, respectively. The signal seed energy matched that of the experiment at 0.7 nJ and the pump energy was varied over the range used in the experiment. 

In order to produce the data shown in Fig. \ref{fig:pumpDepletion}, the field intensities were integrated over the temporal dimension for each spatial grid point. The data shown in Fig. \ref{fig:experimentalData} were generated by additionally integrating over both spatial dimensions. 

\subsection*{Calculation of Non-collinear Phase-Matching Angles}
Non-collinear phase matching data reported in Fig. \ref{fig:NCPM}, \textit{bottom}, were determined from Sellmeier equations (LiNbO3: \cite{Zelmon:97}; BBO: \cite{Tamosauskas:18}; DKDP: \cite{Kirby:87}; CSP: \cite{ZAWILSKI20101127}), using the approach of Ref. \cite{Dean2021Apr}.

\section*{Extended Data}
\setcounter{figure}{0}  

\begin{figure}
    \centering
		\includegraphics[width=\textwidth]{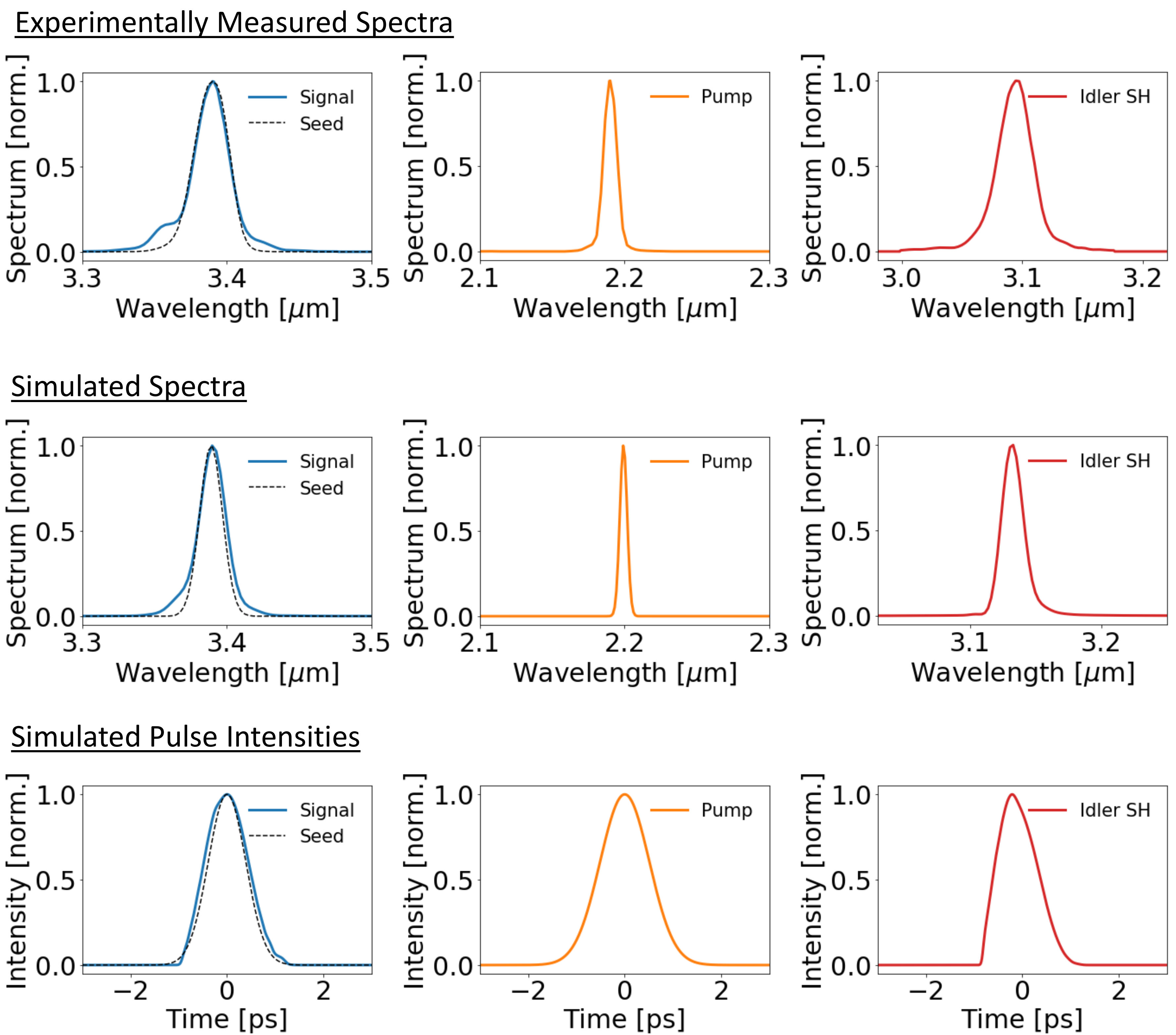}
        \caption{\textbf{(Extended Data)} \textit{Top:} Spectral measurements for the incident seed and pump (at 1W power), and the emitted amplified signal and idler SH beams (after 48-dB signal gain), all normalized to 1. Corresponding simulated spectra (\textit{middle}) and temporal intensity profiles (\textit{bottom}).}
    \label{fig:spectra}
\end{figure}

\textbf{Extended Data Figure 1} \textit{Top:} Spectral measurements for the incident seed and pump (at 1W power), and the emitted amplified signal and idler SH beams (after 48-dB signal gain), all normalized to 1. Corresponding simulated spectra (\textit{middle}) and temporal intensity profiles (\textit{bottom}).

\section*{Supplementary Information}

Supplementary Fig. 1, Section 1, and Video 1.

\section*{Declarations}

\subsection*{Funding}

This material is based upon work primarily supported by the National Science Foundation (NSF) under Grant No. ECCS-1944653. Additional support is acknowledged for the development of specific experimental infrastructure used for this work: by the U.S. Department of Energy (DOE), Office of Science, Basic Energy Sciences (BES), under Award \#DE-SC0020141 (seed laser infrastructure and pulse diagnostics), by the Department of the Navy, Office of Naval Research under Grant No. N00014-19-1-2592 (pump laser development), and by the Cornell Center for Materials Research with funding from the NSF MRSEC program (DMR-1719875) (pump laser and propagation simulation code development).

\subsection*{Authors' contributions}

J.M. supervised the study. N.F. and J.M. conceived and planned the experimental and numerical studies. N.F. performed the numerical analysis. N.F. performed the measurements and analysed the data with assistance from C.D. N.F. and J.Z. developed the pump laser source. D.H. and N.F. developed the seed laser source. D.J.D. calculated noncollinear phase matching angles. K.Z. and P.G.S. manufactured the CSP crystal. All the authors discussed the results and contributed to writing the manuscript.

\bigskip

\begin{appendices}

\end{appendices}


\begin{thebibliography}{10}
\expandafter\ifx\csname url\endcsname\relax
  \def\url#1{\burl{#1}}\fi
\expandafter\ifx\csname urlprefix\endcsname\relax\def\urlprefix{URL }\fi
\providecommand{\bibinfo}[2]{#2}
\providecommand{\eprint}[2][]{\url{#2}}
\providecommand{\doi}[1]{\url{https://doi.org/#1}}
\bibcommenthead

\bibitem{akhmanov1965}
\bibinfo{author}{Akhmanov, S.~A.}, \bibinfo{author}{Kovrigin, A.},
  \bibinfo{author}{Piskarskas, A.}, \bibinfo{author}{Fadeev, V.} \&
  \bibinfo{author}{Khokhlov, R.}
\newblock \bibinfo{title}{Observation of parametric amplification in the
  optical range}.
\newblock \emph{\bibinfo{journal}{Jetp Lett}} \textbf{\bibinfo{volume}{2}}~(7),
  \bibinfo{pages}{191} (\bibinfo{year}{1965}) .

\bibitem{Baumgartner79}
\bibinfo{author}{Baumgartner, R.} \& \bibinfo{author}{Byer, R.}
\newblock \bibinfo{title}{Optical parametric amplification}.
\newblock \emph{\bibinfo{journal}{IEEE Journal of Quantum Electronics}}
  \textbf{\bibinfo{volume}{15}}~(6), \bibinfo{pages}{432--444}
  (\bibinfo{year}{1979}).
\newblock \doi{10.1109/JQE.1979.1070043} .

\bibitem{DUBIETIS92}
\bibinfo{author}{Dubietis, A.}, \bibinfo{author}{Jonušauskas, G.} \&
  \bibinfo{author}{Piskarskas, A.}
\newblock \bibinfo{title}{Powerful femtosecond pulse generation by chirped and
  stretched pulse parametric amplification in {BBO} crystal}.
\newblock \emph{\bibinfo{journal}{Optics Communications}}
  \textbf{\bibinfo{volume}{88}}~(4), \bibinfo{pages}{437--440}
  (\bibinfo{year}{1992}).
\newblock
  \urlprefix\url{https://www.sciencedirect.com/science/article/pii/0030401892900708}.
\newblock \doi{https://doi.org/10.1016/0030-4018(92)90070-8} .

\bibitem{Cerullo03}
\bibinfo{author}{Cerullo, G.} \& \bibinfo{author}{De~Silvestri, S.}
\newblock \bibinfo{title}{Ultrafast optical parametric amplifiers}.
\newblock \emph{\bibinfo{journal}{Review of Scientific Instruments}}
  \textbf{\bibinfo{volume}{74}}~(1), \bibinfo{pages}{1--18}
  (\bibinfo{year}{2003}).
\newblock \urlprefix\url{https://doi.org/10.1063/1.1523642}.
\newblock \doi{10.1063/1.1523642},
  \bibinfo{eprint}{{\href{https://arxiv.org/abs/https://doi.org/10.1063/1.1523642}{{https://doi.org/10.1063/1.1523642}}}}
  .

\bibitem{Fattahi2014}
\bibinfo{author}{Fattahi, H.} \emph{et~al.}
\newblock \bibinfo{title}{Third-generation femtosecond technology}.
\newblock \emph{\bibinfo{journal}{Optica}} \textbf{\bibinfo{volume}{1}}~(1),
  \bibinfo{pages}{45--63} (\bibinfo{year}{2014}).
\newblock
  \urlprefix\url{http://www.osapublishing.org/optica/abstract.cfm?URI=optica-1-1-45}.
\newblock \doi{10.1364/OPTICA.1.000045} .

\bibitem{Baltuska02}
\bibinfo{author}{Baltu\ifmmode~\check{s}\else \v{s}\fi{}ka, A.},
  \bibinfo{author}{Fuji, T.} \& \bibinfo{author}{Kobayashi, T.}
\newblock \bibinfo{title}{Controlling the carrier-envelope phase of ultrashort
  light pulses with optical parametric amplifiers}.
\newblock \emph{\bibinfo{journal}{Phys. Rev. Lett.}}
  \textbf{\bibinfo{volume}{88}}, \bibinfo{pages}{133901}
  (\bibinfo{year}{2002}).
\newblock
  \urlprefix\url{https://link.aps.org/doi/10.1103/PhysRevLett.88.133901}.
\newblock \doi{10.1103/PhysRevLett.88.133901} .

\bibitem{Ledingham03}
\bibinfo{author}{Ledingham, K. W.~D.}, \bibinfo{author}{McKenna, P.} \&
  \bibinfo{author}{Singhal, R.~P.}
\newblock \bibinfo{title}{Applications for nuclear phenomena generated by
  ultra-intense lasers}.
\newblock \emph{\bibinfo{journal}{Science}}
  \textbf{\bibinfo{volume}{300}}~(5622), \bibinfo{pages}{1107--1111}
  (\bibinfo{year}{2003}).
\newblock
  \urlprefix\url{https://www.science.org/doi/abs/10.1126/science.1080552}.
\newblock \doi{10.1126/science.1080552},
  \bibinfo{eprint}{{\href{https://arxiv.org/abs/https://www.science.org/doi/pdf/10.1126/science.1080552}{{https://www.science.org/doi/pdf/10.1126/science.1080552}}}}
  .

\bibitem{Nibbering05}
\bibinfo{author}{Nibbering, E.~T.}, \bibinfo{author}{Fidder, H.} \&
  \bibinfo{author}{Pines, E.}
\newblock \bibinfo{title}{Ultrafast chemistry: Using time-resolved vibrational
  spectroscopy for interrogation of structural dynamics}.
\newblock \emph{\bibinfo{journal}{Annual Review of Physical Chemistry}}
  \textbf{\bibinfo{volume}{56}}~(1), \bibinfo{pages}{337--367}
  (\bibinfo{year}{2005}).
\newblock
  \urlprefix\url{https://doi.org/10.1146/annurev.physchem.56.092503.141314}.
\newblock \doi{10.1146/annurev.physchem.56.092503.141314}, \bibinfo{note}{pMID:
  15796704},
  \bibinfo{eprint}{{\href{https://arxiv.org/abs/https://doi.org/10.1146/annurev.physchem.56.092503.141314}{{https://doi.org/10.1146/annurev.physchem.56.092503.141314}}}}
  .

\bibitem{Mourou2006}
\bibinfo{author}{Mourou, G.~A.}, \bibinfo{author}{Tajima, T.} \&
  \bibinfo{author}{Bulanov, S.~V.}
\newblock \bibinfo{title}{Optics in the relativistic regime}.
\newblock \emph{\bibinfo{journal}{Rev. Mod. Phys.}}
  \textbf{\bibinfo{volume}{78}}, \bibinfo{pages}{309--371}
  (\bibinfo{year}{2006}).
\newblock \urlprefix\url{https://link.aps.org/doi/10.1103/RevModPhys.78.309}.
\newblock \doi{10.1103/RevModPhys.78.309} .

\bibitem{SALAMIN2006}
\bibinfo{author}{Salamin, Y.~I.}, \bibinfo{author}{Hu, S.},
  \bibinfo{author}{Hatsagortsyan, K.~Z.} \& \bibinfo{author}{Keitel, C.~H.}
\newblock \bibinfo{title}{Relativistic high-power laser–matter interactions}.
\newblock \emph{\bibinfo{journal}{Physics Reports}}
  \textbf{\bibinfo{volume}{427}}~(2), \bibinfo{pages}{41--155}
  (\bibinfo{year}{2006}).
\newblock
  \urlprefix\url{https://www.sciencedirect.com/science/article/pii/S0370157306000093}.
\newblock \doi{https://doi.org/10.1016/j.physrep.2006.01.002} .

\bibitem{Sansone2011}
\bibinfo{author}{Sansone, G.}, \bibinfo{author}{Poletto, L.} \&
  \bibinfo{author}{Nisoli, M.}
\newblock \bibinfo{title}{High-energy attosecond light sources}.
\newblock \emph{\bibinfo{journal}{Nature Photonics}}
  \textbf{\bibinfo{volume}{5}}~(11), \bibinfo{pages}{655--663}
  (\bibinfo{year}{2011}).
\newblock \urlprefix\url{https://doi.org/10.1038/nphoton.2011.167}.
\newblock \doi{10.1038/nphoton.2011.167} .

\bibitem{Danson15}
\bibinfo{author}{Danson, C.}, \bibinfo{author}{Hillier, D.},
  \bibinfo{author}{Hopps, N.} \& \bibinfo{author}{Neely, D.}
\newblock \bibinfo{title}{Petawatt class lasers worldwide}.
\newblock \emph{\bibinfo{journal}{High Power Laser Science and Engineering}}
  \textbf{\bibinfo{volume}{3}}, \bibinfo{pages}{e3} (\bibinfo{year}{2015}).
\newblock \doi{10.1017/hpl.2014.52} .

\bibitem{Armstrong:62}
\bibinfo{author}{Armstrong, J.~A.}, \bibinfo{author}{Bloembergen, N.},
  \bibinfo{author}{Ducuing, J.} \& \bibinfo{author}{Pershan, P.~S.}
\newblock \bibinfo{title}{Interactions between light waves in a nonlinear
  dielectric}.
\newblock \emph{\bibinfo{journal}{Phys. Rev.}} \textbf{\bibinfo{volume}{127}},
  \bibinfo{pages}{1918--1939} (\bibinfo{year}{1962}).
\newblock \urlprefix\url{https://link.aps.org/doi/10.1103/PhysRev.127.1918}.
\newblock \doi{10.1103/PhysRev.127.1918} .

\bibitem{Jauregui2013}
\bibinfo{author}{Jauregui, C.}, \bibinfo{author}{Limpert, J.} \&
  \bibinfo{author}{T{\"u}nnermann, A.}
\newblock \bibinfo{title}{High-power fibre lasers}.
\newblock \emph{\bibinfo{journal}{Nature Photonics}}
  \textbf{\bibinfo{volume}{7}}~(11), \bibinfo{pages}{861--867}
  (\bibinfo{year}{2013}).
\newblock \urlprefix\url{https://doi.org/10.1038/nphoton.2013.273}.
\newblock \doi{10.1038/nphoton.2013.273} .

\bibitem{Sincore18}
\bibinfo{author}{Sincore, A.}, \bibinfo{author}{Bradford, J.~D.},
  \bibinfo{author}{Cook, J.}, \bibinfo{author}{Shah, L.} \&
  \bibinfo{author}{Richardson, M.~C.}
\newblock \bibinfo{title}{High average power thulium-doped silica fiber lasers:
  Review of systems and concepts}.
\newblock \emph{\bibinfo{journal}{IEEE Journal of Selected Topics in Quantum
  Electronics}} \textbf{\bibinfo{volume}{24}}~(3), \bibinfo{pages}{1--8}
  (\bibinfo{year}{2018}).
\newblock \doi{10.1109/JSTQE.2017.2775964} .

\bibitem{Saraceno2019}
\bibinfo{author}{Saraceno, C.~J.}, \bibinfo{author}{Sutter, D.},
  \bibinfo{author}{Metzger, T.} \& \bibinfo{author}{Abdou~Ahmed, M.}
\newblock \bibinfo{title}{The amazing progress of high-power ultrafast
  thin-disk lasers}.
\newblock \emph{\bibinfo{journal}{Journal of the European Optical Society-Rapid
  Publications}} \textbf{\bibinfo{volume}{15}}~(1), \bibinfo{pages}{15}
  (\bibinfo{year}{2019}).
\newblock \urlprefix\url{https://doi.org/10.1186/s41476-019-0108-1}.
\newblock \doi{10.1186/s41476-019-0108-1} .

\bibitem{Muller:20}
\bibinfo{author}{M\"{u}ller, M.} \emph{et~al.}
\newblock \bibinfo{title}{10.4kw coherently combined ultrafast fiber
  laser}.
\newblock \emph{\bibinfo{journal}{Opt. Lett.}}
  \textbf{\bibinfo{volume}{45}}~(11), \bibinfo{pages}{3083--3086}
  (\bibinfo{year}{2020}).
\newblock
  \urlprefix\url{http://opg.optica.org/ol/abstract.cfm?URI=ol-45-11-3083}.
\newblock \doi{10.1364/OL.392843} .

\bibitem{Bagnoud:05}
\bibinfo{author}{Bagnoud, V.}, \bibinfo{author}{Begishev, I.},
  \bibinfo{author}{Guardalben, M.}, \bibinfo{author}{Puth, J.} \&
  \bibinfo{author}{Zuegel, J.}
\newblock \bibinfo{title}{5 hz, $>$ 250 {mJ} optical parametric chirped-pulse
  amplifier at 1053 nm}.
\newblock \emph{\bibinfo{journal}{Opt. Lett.}} \textbf{\bibinfo{volume}{30}},
  \bibinfo{pages}{1843--1845} (\bibinfo{year}{2005}) .

\bibitem{Begishev:90}
\bibinfo{author}{Begishev, I.~A.} \emph{et~al.}
\newblock \bibinfo{title}{Highly efficient parametric amplification of optical
  beams. i. optimization of the profiles of interacting waves in parametric
  amplification}.
\newblock \emph{\bibinfo{journal}{Sov. J. Quantum Electron.}}
  \textbf{\bibinfo{volume}{20}}~(9), \bibinfo{pages}{1100--1103}
  (\bibinfo{year}{1990}).
\newblock \urlprefix\url{https://doi.org/10.1070%2Fqe1990v020n09abeh007413}.
\newblock \doi{10.1070/qe1990v020n09abeh007413} .

\bibitem{Moses:11}
\bibinfo{author}{Moses, J.} \& \bibinfo{author}{Huang, S.-W.}
\newblock \bibinfo{title}{Conformal profile theory for performance scaling of
  ultrabroadband optical parametric chirped pulse amplification}.
\newblock \emph{\bibinfo{journal}{J. Opt. Soc. Am. B}}
  \textbf{\bibinfo{volume}{28}}, \bibinfo{pages}{812--831}
  (\bibinfo{year}{2011}) .

\bibitem{Cao:18}
\bibinfo{author}{Cao, H.}, \bibinfo{author}{T{\'o}th, S.},
  \bibinfo{author}{Kalashnikov, M.}, \bibinfo{author}{Chvykov, V.} \&
  \bibinfo{author}{Osvay, K.}
\newblock \bibinfo{title}{Highly efficient, cascaded extraction optical
  parametric amplifier}.
\newblock \emph{\bibinfo{journal}{Opt Express}} \textbf{\bibinfo{volume}{26}},
  \bibinfo{pages}{7516--7527} (\bibinfo{year}{2018}) .

\bibitem{Mackonis:20}
\bibinfo{author}{Mackonis, P.} \& \bibinfo{author}{Rodin, A.~M.}
\newblock \bibinfo{title}{{OPCPA} investigation with control over the temporal
  shape of 1.2 ps pump pulses}.
\newblock \emph{\bibinfo{journal}{Opt. Express}}
  \textbf{\bibinfo{volume}{28}}~(8), \bibinfo{pages}{12020--12027}
  (\bibinfo{year}{2020}).
\newblock
  \urlprefix\url{http://opg.optica.org/oe/abstract.cfm?URI=oe-28-8-12020}.
\newblock \doi{10.1364/OE.383754} .

\bibitem{Fischer:21}
\bibinfo{author}{Fischer, P.}, \bibinfo{author}{Muschet, A.},
  \bibinfo{author}{Lang, T.}, \bibinfo{author}{Salh, R.} \&
  \bibinfo{author}{Veisz, L.}
\newblock \bibinfo{title}{Saturation control of an optical parametric
  chirped-pulse amplifier}.
\newblock \emph{\bibinfo{journal}{Opt. Express}}
  \textbf{\bibinfo{volume}{29}}~(3), \bibinfo{pages}{4210--4218}
  (\bibinfo{year}{2021}).
\newblock
  \urlprefix\url{http://opg.optica.org/oe/abstract.cfm?URI=oe-29-3-4210}.
\newblock \doi{10.1364/OE.415564} .

\bibitem{Jeys:96}
\bibinfo{author}{Jeys, T.~H.}
\newblock \bibinfo{title}{Multipass optical parametric amplifier}.
\newblock \emph{\bibinfo{journal}{Opt. Lett.}}
  \textbf{\bibinfo{volume}{21}}~(16), \bibinfo{pages}{1229--1231}
  (\bibinfo{year}{1996}).
\newblock
  \urlprefix\url{http://opg.optica.org/ol/abstract.cfm?URI=ol-21-16-1229}.
\newblock \doi{10.1364/OL.21.001229} .

\bibitem{Ma:15}
\bibinfo{author}{Ma, J.} \emph{et~al.}
\newblock \bibinfo{title}{Quasi-parametric amplification of chirped pulses
  based on a {Sm\textsuperscript{3+}}-doped yttrium calcium oxyborate crystal}.
\newblock \emph{\bibinfo{journal}{Optica}} \textbf{\bibinfo{volume}{2}},
  \bibinfo{pages}{1006--1009} (\bibinfo{year}{2015}) .

\bibitem{El-Ganainy:15}
\bibinfo{author}{El-Ganainy, R.}, \bibinfo{author}{Dadap, J.~I.} \&
  \bibinfo{author}{Osgood, R.~M.}
\newblock \bibinfo{title}{Optical parametric amplification via non-hermitian
  phase matching}.
\newblock \emph{\bibinfo{journal}{Opt. Lett.}}
  \textbf{\bibinfo{volume}{40}}~(21), \bibinfo{pages}{5086--5089}
  (\bibinfo{year}{2015}).
\newblock \urlprefix\url{http://ol.osa.org/abstract.cfm?URI=ol-40-21-5086}.
\newblock \doi{10.1364/OL.40.005086} .

\bibitem{Zhong:16}
\bibinfo{author}{Zhong, Q.}, \bibinfo{author}{Ahmed, A.},
  \bibinfo{author}{Dadap, J.~I.}, \bibinfo{author}{Osgood, R.~M.} \&
  \bibinfo{author}{El-Ganainy, R.}
\newblock \bibinfo{title}{Parametric amplification in quasi-{PT} symmetric
  coupled waveguide structures}.
\newblock \emph{\bibinfo{journal}{New J. Phys.}} \textbf{\bibinfo{volume}{18}},
  \bibinfo{pages}{125006} (\bibinfo{year}{2016}) .

\bibitem{Ma:17}
\bibinfo{author}{Ma, J.} \emph{et~al.}
\newblock \bibinfo{title}{Broadband, efficient, and robust quasi-parametric
  chirped-pulse amplification}.
\newblock \emph{\bibinfo{journal}{Opt. Express}}
  \textbf{\bibinfo{volume}{25}}~(21), \bibinfo{pages}{25149--25164}
  (\bibinfo{year}{2017}).
\newblock
  \urlprefix\url{http://www.opticsexpress.org/abstract.cfm?URI=oe-25-21-25149}.
\newblock \doi{10.1364/OE.25.025149} .

\bibitem{Perego2018}
\bibinfo{author}{Perego, A.~M.}, \bibinfo{author}{Turitsyn, S.~K.} \&
  \bibinfo{author}{Staliunas, K.}
\newblock \bibinfo{title}{Gain through losses in nonlinear optics}.
\newblock \emph{\bibinfo{journal}{Light: Science {\&} Applications}}
  \textbf{\bibinfo{volume}{7}}~(1), \bibinfo{pages}{43} (\bibinfo{year}{2018}).
\newblock \urlprefix\url{https://doi.org/10.1038/s41377-018-0042-9}.
\newblock \doi{10.1038/s41377-018-0042-9} .

\bibitem{Feng2017}
\bibinfo{author}{Feng, L.}, \bibinfo{author}{El-Ganainy, R.} \&
  \bibinfo{author}{Ge, L.}
\newblock \bibinfo{title}{Non-hermitian photonics based on parity--time
  symmetry}.
\newblock \emph{\bibinfo{journal}{Nature Photonics}}
  \textbf{\bibinfo{volume}{11}}~(12), \bibinfo{pages}{752--762}
  (\bibinfo{year}{2017}).
\newblock \urlprefix\url{https://doi.org/10.1038/s41566-017-0031-1}.
\newblock \doi{10.1038/s41566-017-0031-1} .

\bibitem{Flemens:21}
\bibinfo{author}{Flemens, N.}, \bibinfo{author}{Swenson, N.} \&
  \bibinfo{author}{Moses, J.}
\newblock \bibinfo{title}{Efficient parametric amplification via simultaneous
  second harmonic generation}.
\newblock \emph{\bibinfo{journal}{Opt. Express}}
  \textbf{\bibinfo{volume}{29}}~(19), \bibinfo{pages}{30590--30609}
  (\bibinfo{year}{2021}).
\newblock
  \urlprefix\url{http://www.osapublishing.org/oe/abstract.cfm?URI=oe-29-19-30590}.
\newblock \doi{10.1364/OE.437864} .

\bibitem{Sanchez:14}
\bibinfo{author}{S\'{a}nchez, D.} \emph{et~al.}
\newblock \bibinfo{title}{Broadband mid-{IR} frequency comb with {CdSiP}$_2$
  and {AgGaS}$_2$ from an {Er,Tm:Ho} fiber laser}.
\newblock \emph{\bibinfo{journal}{Opt. Lett.}}
  \textbf{\bibinfo{volume}{39}}~(24), \bibinfo{pages}{6883--6886}
  (\bibinfo{year}{2014}).
\newblock \urlprefix\url{http://ol.osa.org/abstract.cfm?URI=ol-39-24-6883}.
\newblock \doi{10.1364/OL.39.006883} .

\bibitem{Liang:17}
\bibinfo{author}{Liang, H.} \emph{et~al.}
\newblock \bibinfo{title}{High-energy mid-infrared sub-cycle pulse synthesis
  from a parametric amplifier}.
\newblock \emph{\bibinfo{journal}{Nat. Commun.}} \textbf{\bibinfo{volume}{8}},
  \bibinfo{pages}{141} (\bibinfo{year}{2017}) .

\bibitem{Dean2021Apr}
\bibinfo{author}{Dean, D.~J.}, \bibinfo{author}{Flemens, N.},
  \bibinfo{author}{Heberle, D.} \& \bibinfo{author}{Moses, J.}
\newblock \bibinfo{title}{ in \textit{Widely tunable second harmonic
  amplification by noncollinear phase matching in bulk birefringent materials}}
  , Vol. \bibinfo{volume}{11670} \bibinfo{pages}{70--78}
  (\bibinfo{publisher}{SPIE}, \bibinfo{year}{2021}).

\bibitem{Manzoni:11}
\bibinfo{author}{Manzoni, C.}, \bibinfo{author}{Moses, J.},
  \bibinfo{author}{K\"{a}rtner, F.~X.} \& \bibinfo{author}{Cerullo, G.}
\newblock \bibinfo{title}{Excess quantum noise in optical parametric
  chirped-pulse amplification}.
\newblock \emph{\bibinfo{journal}{Opt. Express}}
  \textbf{\bibinfo{volume}{19}}~(9), \bibinfo{pages}{8357--8366}
  (\bibinfo{year}{2011}).
\newblock
  \urlprefix\url{http://opg.optica.org/oe/abstract.cfm?URI=oe-19-9-8357}.
\newblock \doi{10.1364/OE.19.008357} .

\bibitem{Feng:21}
\bibinfo{author}{Feng, C.} \emph{et~al.}
\newblock \bibinfo{title}{Analysis of pump-to-signal noise transfer in
  two-stage ultra-broadband optical parametric chirped-pulse amplification}.
\newblock \emph{\bibinfo{journal}{Opt. Express}}
  \textbf{\bibinfo{volume}{29}}~(24), \bibinfo{pages}{40240--40258}
  (\bibinfo{year}{2021}).
\newblock
  \urlprefix\url{http://opg.optica.org/oe/abstract.cfm?URI=oe-29-24-40240}.
\newblock \doi{10.1364/OE.441108} .

\bibitem{Hu2020}
\bibinfo{author}{Hu, D.} \emph{et~al.}
\newblock \bibinfo{title}{Comparative study on coherent noise in optical
  parametric and quasi-parametric chirped-pulse amplification}.
\newblock \emph{\bibinfo{journal}{Optics Communications}}
  \textbf{\bibinfo{volume}{464}}, \bibinfo{pages}{125461}
  (\bibinfo{year}{2020}).
\newblock
  \urlprefix\url{https://www.sciencedirect.com/science/article/pii/S0030401820301164}.
\newblock \doi{https://doi.org/10.1016/j.optcom.2020.125461} .

\bibitem{Luo:06}
\bibinfo{author}{Luo, H.}, \bibinfo{author}{Qian, L.}, \bibinfo{author}{Yuan,
  P.} \& \bibinfo{author}{Zhu, H.}
\newblock \bibinfo{title}{Generation of tunable narrowband pulses initiating
  from a femtosecond optical parametric amplifier}.
\newblock \emph{\bibinfo{journal}{Opt. Express}}
  \textbf{\bibinfo{volume}{14}}~(22), \bibinfo{pages}{10631--10635}
  (\bibinfo{year}{2006}).
\newblock
  \urlprefix\url{http://opg.optica.org/oe/abstract.cfm?URI=oe-14-22-10631}.
\newblock \doi{10.1364/OE.14.010631} .

\bibitem{Schmidt2014}
\bibinfo{author}{Schmidt, B.~E.} \emph{et~al.}
\newblock \bibinfo{title}{Frequency domain optical parametric amplification}.
\newblock \emph{\bibinfo{journal}{Nature Communications}}
  \textbf{\bibinfo{volume}{5}}~(1), \bibinfo{pages}{3643}
  (\bibinfo{year}{2014}).
\newblock \urlprefix\url{https://doi.org/10.1038/ncomms4643}.
\newblock \doi{10.1038/ncomms4643} .

\bibitem{Leblanc:20}
\bibinfo{author}{Leblanc, A.} \emph{et~al.}
\newblock \bibinfo{title}{High-field mid-infrared pulses derived from frequency
  domain optical parametric amplification}.
\newblock \emph{\bibinfo{journal}{Opt. Lett.}}
  \textbf{\bibinfo{volume}{45}}~(8), \bibinfo{pages}{2267--2270}
  (\bibinfo{year}{2020}).
\newblock
  \urlprefix\url{http://opg.optica.org/ol/abstract.cfm?URI=ol-45-8-2267}.
\newblock \doi{10.1364/OL.389804} .

\bibitem{Chou:99}
\bibinfo{author}{Chou, M.~H.}, \bibinfo{author}{Parameswaran, K.~R.},
  \bibinfo{author}{Fejer, M.~M.} \& \bibinfo{author}{Brener, I.}
\newblock \bibinfo{title}{Multiple-channel wavelength conversion by use of
  engineered quasi-phase-matching structures in {LiNbO\textsubscript{3}}
  waveguides}.
\newblock \emph{\bibinfo{journal}{Opt. Lett.}}
  \textbf{\bibinfo{volume}{24}}~(16), \bibinfo{pages}{1157--1159}
  (\bibinfo{year}{1999}).
\newblock \urlprefix\url{http://ol.osa.org/abstract.cfm?URI=ol-24-16-1157}.
\newblock \doi{10.1364/OL.24.001157} .

\bibitem{Liu:02}
\bibinfo{author}{Liu, Z.-W.} \emph{et~al.}
\newblock \bibinfo{title}{Engineering of a dual-periodic optical superlattice
  used in a coupled optical parametric interaction}.
\newblock \emph{\bibinfo{journal}{J. Opt. Soc. Am. B}}
  \textbf{\bibinfo{volume}{19}}~(7), \bibinfo{pages}{1676--1684}
  (\bibinfo{year}{2002}).
\newblock
  \urlprefix\url{http://josab.osa.org/abstract.cfm?URI=josab-19-7-1676}.
\newblock \doi{10.1364/JOSAB.19.001676} .

\bibitem{Norton:03}
\bibinfo{author}{Norton, A.~H.} \& \bibinfo{author}{de~Sterke, C.~M.}
\newblock \bibinfo{title}{Optimal poling of nonlinear photonic crystals for
  frequency conversion}.
\newblock \emph{\bibinfo{journal}{Opt Lett.}} \textbf{\bibinfo{volume}{28}},
  \bibinfo{pages}{188--190} (\bibinfo{year}{2003}) .

\bibitem{Lifshitz:05}
\bibinfo{author}{R.~Lifshitz, A.~B., A.~Arie}.
\newblock \bibinfo{title}{Photonic quasicrystals for nonlinear optical
  frequency conversion}.
\newblock \emph{\bibinfo{journal}{Phys. Rev. Lett.}}
  \textbf{\bibinfo{volume}{95}}, \bibinfo{pages}{133901}
  (\bibinfo{year}{2005}).
\newblock
  \urlprefix\url{https://link.aps.org/doi/10.1103/PhysRevLett.95.133901}.
\newblock \doi{10.1103/PhysRevLett.95.133901} .

\bibitem{Krogen2017Apr}
\bibinfo{author}{Krogen, P.} \emph{et~al.}
\newblock \bibinfo{title}{{Generation and multi-octave shaping of mid-infrared
  intense single-cycle pulses}}.
\newblock \emph{\bibinfo{journal}{Nat. Photonics}}
  \textbf{\bibinfo{volume}{11}}~(4), \bibinfo{pages}{222--226}
  (\bibinfo{year}{2017}).
\newblock \doi{10.1038/nphoton.2017.34} .

\bibitem{ZAWILSKI20101127}
\bibinfo{author}{Zawilski, K.~T.} \emph{et~al.}
\newblock \bibinfo{title}{Growth and characterization of large
  {CdSiP\textsubscript{2}} single crystals}.
\newblock \emph{\bibinfo{journal}{Journal of Crystal Growth}}
  \textbf{\bibinfo{volume}{312}}~(8), \bibinfo{pages}{1127--1132}
  (\bibinfo{year}{2010}).
\newblock
  \urlprefix\url{https://www.sciencedirect.com/science/article/pii/S0022024809009427}.
\newblock \doi{https://doi.org/10.1016/j.jcrysgro.2009.10.034},
  \bibinfo{note}{the 17th American Conference on Crystal Growth and Epitaxy/The
  14th US Biennial Workshop on Organometallic Vapor Phase Epitaxy/The 6th
  International Workshop on Modeling in Crystal Growth} .

\bibitem{Zelmon:97}
\bibinfo{author}{Zelmon, D.~E.}, \bibinfo{author}{Small, D.~L.} \&
  \bibinfo{author}{Jundt, D.}
\newblock \bibinfo{title}{Infrared corrected sellmeier coefficients for
  congruently grown lithium niobate and 5 mol.\% magnesium oxide--doped
  lithium niobate}.
\newblock \emph{\bibinfo{journal}{J. Opt. Soc. Am. B}}
  \textbf{\bibinfo{volume}{14}}~(12), \bibinfo{pages}{3319--3322}
  (\bibinfo{year}{1997}).
\newblock
  \urlprefix\url{http://opg.optica.org/josab/abstract.cfm?URI=josab-14-12-3319}.
\newblock \doi{10.1364/JOSAB.14.003319} .

\bibitem{Tamosauskas:18}
\bibinfo{author}{Tamo\v{s}auskas, G.}, \bibinfo{author}{Beresnevi\v{c}ius, G.},
  \bibinfo{author}{Gadonas, D.} \& \bibinfo{author}{Dubietis, A.}
\newblock \bibinfo{title}{Transmittance and phase matching of bbo crystal in
  the 3-5 \textmu m range and its application for the characterization of
  mid-infrared laser pulses}.
\newblock \emph{\bibinfo{journal}{Opt. Mater. Express}}
  \textbf{\bibinfo{volume}{8}}~(6), \bibinfo{pages}{1410--1418}
  (\bibinfo{year}{2018}).
\newblock
  \urlprefix\url{http://opg.optica.org/ome/abstract.cfm?URI=ome-8-6-1410}.
\newblock \doi{10.1364/OME.8.001410} .

\bibitem{Kirby:87}
\bibinfo{author}{Kirby, K.~W.} \& \bibinfo{author}{DeShazer, L.~G.}
\newblock \bibinfo{title}{Refractive indices of 14 nonlinear crystals
  isomorphic to {KH\textsubscript{2}PO\textsubscript{4}}}.
\newblock \emph{\bibinfo{journal}{J. Opt. Soc. Am. B}}
  \textbf{\bibinfo{volume}{4}}~(7), \bibinfo{pages}{1072--1078}
  (\bibinfo{year}{1987}).
\newblock
  \urlprefix\url{http://opg.optica.org/josab/abstract.cfm?URI=josab-4-7-1072}.
\newblock \doi{10.1364/JOSAB.4.001072} .

\end{thebibliography}
\end{document}

\title[Supplementary Information]{Supplementary Information for High Quantum Efficiency Parametric Amplification via Hybridized Nonlinear Optics}

\author*[1]{\fnm{Noah} \sur{Flemens}}\email{nrf33@cornell.edu}

\author[1]{\fnm{Dylan} \sur{Heberle}}

\author[1]{\fnm{Jiaoyang} \sur{Zheng}}

\author[1]{\fnm{Devin J.} \sur{Dean}}

\author[1]{\fnm{Connor} \sur{Davis}}

\author[2]{\fnm{Kevin} \sur{Zawilski}}

\author[2]{\fnm{Peter G.} \sur{Schunemann}}

\author*[1]{\fnm{Jeffrey} \sur{Moses}}\email{moses@cornell.edu}

\affil[1]{\orgdiv{School of Applied and Engineering Physics}, \orgname{Cornell University}, \orgaddress{\street{142 Sciences Drive}, \city{Ithaca}, \postcode{14853}, \state{New York}, \country{United States}}}

\affil[2]{\orgname{BAE Systems}, \orgaddress{\street{P.O. Box 868, MER15-1813}, \city{Nashua}, \postcode{03061}, \state{New Hampshire}, \country{United States}}}

\maketitle

\newpage

\subsection*{1. Alternate determination of amplifier quantum efficiency from pump depletion measurements}

We independently verified the quantum efficiency measurements by measuring the pump depletion efficiency. Since each pump photon produces exactly one signal photon, the internal fractional pump photon depletion should exactly match the quantum efficiency determined from signal energy measurements. To determine the internal pump depletion, we directly measured and took the ratio of the pump power emitted from the amplifier crystal after a 2250$\pm$500 nm bandpass filter to the measured incident pump power, taking into consideration the 4\% refelection loss in each case. There was no discernible change in pump power transmission through the crystal due to superfluorescence when the seed was blocked. We independently verified that the ratio of incident and emitted pump power for the unseeded amplifier did not depend on the pump power, and found it consistent with 4\% losses due to reflection at each crystal face. Thus, we determined there were no significant nonlinear processes occurring for the unseeded amplifier that could affect the emitted pump intensity.

The experimentally measured percent unconverted pump is plotted in Fig. \ref{fig:residualpump} against the expected values from simulations of HPA and OPA, and we observe the same close match between measured and simulated HPA data illustrated in Fig. 3 of the main text. The pump depletion with the full input pump power is 32\%, matching the 68\% quantum efficiency calculated from the signal efficiency measurement.

\setcounter{figure}{0}  

\begin{figure}
    \centering
	\includegraphics[width=\textwidth]{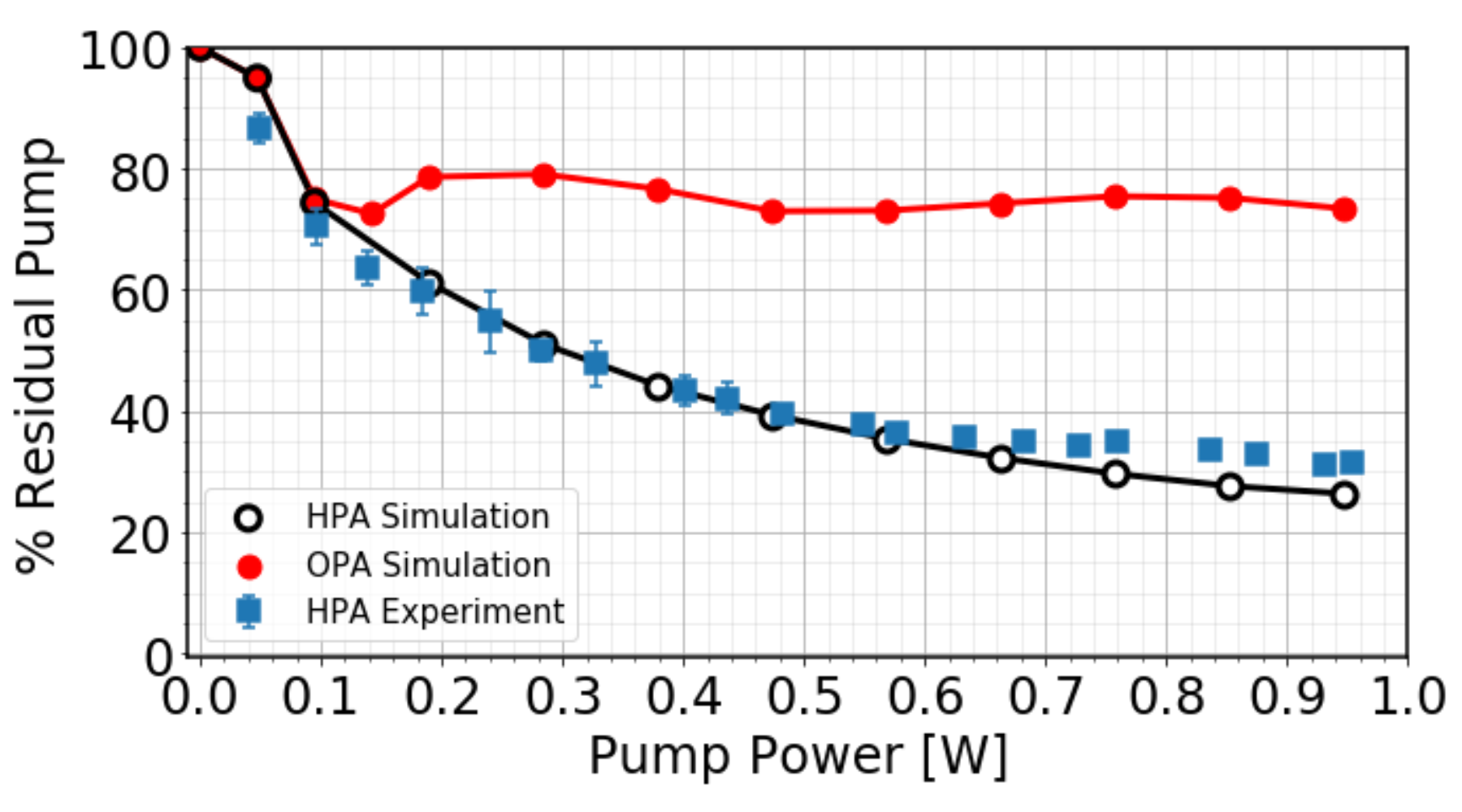}
        \caption{ Percent ratio of output to incident pump power (\% Residual Pump) vs incident pump power, comparing the HPA experiment (blue squares) to the expected pump depletion from simulated HPA (black circles) and the equivalent OPA pump depletion without SHG (red circles).}
    \label{fig:residualpump}
\end{figure}